\input{aipcheck}

\documentclass[
    ,final            
  ]
  {aipproc}

\layoutstyle{6x9}

\begin{document}

\title{Effect of three-pion unitarity on resonance poles from heavy meson
decays}



\author{Satoshi X. Nakamura}{
  address={Excited Baryon Analysis Center, Jefferson Laboratory, Newport News, Virginia 23606, USA}
}

\begin{abstract}
We study the final state interaction in 3$\pi$ decay of meson
resonances at the Excited Baryon Analysis Center (EBAC)
of JLab.
We apply the dynamical coupled-channels formulation which has been
extensively used by EBAC to extract N* information.  
The formulation satisfies the 3$\pi$ unitarity condition which has been
missed in the existing works with the isobar models. 
We report the effect of the 3$\pi$ unitarity on
the meson resonance pole positions and Dalitz plot. 
\end{abstract}

\maketitle


\section{Introduction}

Why do we care the three-pion (or $\pi\pi K$, etc.) scattering system ?
Some of recent and forthcoming experiments would call for serious study
of the system to extract what they expect to find.
One of such experiments aims to find the
so-called exotic mesons
in a reaction like $\pi({\rm or}\ \gamma)N\to M^* N\to \pi\pi\pi N$~\cite{e852,gluex},
where $M^*$ is an intermediate excited meson
that could be exotic.
An exotic meson lies outside of the
constituent quark model and speculated to be a tetraquark state or
hybrid state. 
The spectroscopy of the hybrid states provides information about gluon
self-interactions.
Another experiment seeks for
physics beyond the Standard Model contributing to the CP
violation in B- or D-decays~\cite{babar,belle}.
Both of the experiments involve heavy meson decays at very
short-distance followed by the formation of lighter hadrons
such as pions and/or kaons.
And those light hadrons are to be detected after
a great number of rescatterings.
Thus, for extracting what has happened at the short-distance, it is
essential to understand or precisely model the final state interactions
of the light hadrons.

What has conventionally been used for Dalitz plot analyses
of experiments of this kind is
the so-called isobar model in which it is assumed that two of the three
mesons form an isobar state ($f_0, \rho, K^*$, etc.), and the rest does
not interact with the others (spectator). 
The three-body unitarity is obviously missing, and that is what we will
address here.
(Note that most isobar models do not take care of even two-body
unitarity for the paired mesons.)
Recently, the Excited Baryon Analysis Center of Jefferson Lab 
has extended their dynamical coupled-channels model~\cite{msl}, which has been
extensively used to study baryon resonances, to three-meson scattering~\cite{ebac}.
In this report, we describe our model and then apply it to the
calculation of $\pi_2(1670)$ and $\pi_2(2100)$ pole positions and their
3$\pi$ decays (Dalitz plots).
We will examine how significant 3$\pi$ unitarity is to extract the
properties of these heavy mesons from the Dalitz plots.

\section{Coupled-channels model}
\label{sec_model}

We consider the 3$\pi$ decay of a heavy meson ($M^*\to 3\pi$) and a heavy
meson resonance formed in 3$\pi$ scattering ($3\pi\to M^*\to 3\pi$).
Interesting quantities to calculate are
the decay amplitude (and the corresponding Dalitz plot) for the decay,
and the pole position of the resonance.
In any case, we need 3$\pi$ scattering amplitudes.
For calculating 3$\pi$ amplitude, first we need to develop a more basic
$\pi\pi$ model. In the following, we discuss our $\pi\pi$ model and then
calculation of the 3$\pi$ amplitude.\\

\noindent{\bf $\pi\pi$ model}\ \ 
For a simpler calculation, we employ an isobar-type model for
$\pi\pi$ interaction so that $\pi\pi\to R\to\pi\pi\ ({\rm or}\ K\bar{K})$, 
where $R$ ($= f_0, \rho$ or $f_2$) is an isobar. 
The $\pi\pi$ potentials for a partial wave (orbital
angular momentum $L$, total isospin $I$) are parametrized as
$\sum_{R}f^{LI}_{R,i}(p')(W-m_{R})^{-1}f^{LI}_{R,j}(p)$,
where $W$ and $m_R$ are the total energy and the bare mass of the isobar
$R$, respectively;
$f_{R,i}(p)$ is the $R\leftrightarrow i$ $(i=\pi\pi\ {\rm or}\ K\bar{K})$ vertex with $p$
being the relative momentum of $\pi\pi$ ($K\bar{K}$).
The $\pi\pi$ scattering amplitude 
is obtained by solving the coupled-channels
Lippmann-Schwinger equation with this potential.\\
%
%

\noindent{\bf $\pi$-isobar scattering equation}\ \ 
Because we have introduced the isobars ($R$), 
the partial wave amplitude specified by $J^P$ (total spin and parity) 
and $T$ (total isospin) can be obtained by solving the corresponding 
$\pi-R$ scattering equation:
\begin{eqnarray}
\label{eq:tmat}
T^{J^PT}_{\alpha,\beta} (p',p; W)=V^{J^PT}_{\alpha,\beta}(p',p; W)
+\sum_{\gamma\gamma'}
\int^\infty_0\!\!\! q^2dq\ 
V^{J^PT}_{\alpha,\gamma}(p',q;W)\ 
G_{\gamma\gamma'}(q,W)
T^{J^PT}_{\gamma',\beta}(q,p;W)
\ ,
\end{eqnarray}
where the indices $\alpha,\beta,\gamma$ specify channels
($\pi f_0$ or $\pi \rho$ or $\pi f_2$),
$E_{\gamma}(q)=\sqrt{m_\pi^2+q^2}+\sqrt{m_{R_\gamma}^2+q^2}$.
The $\pi-R$ Green function, $G_{\gamma\gamma'}(q,W)$, is given by
$[G^{-1}(q,W)]_{\gamma\gamma'}=
[W-E_{\gamma}(q)]\delta_{\gamma\gamma'}-\Sigma_{\gamma\gamma'}(q,W)$,
with the self-energy ($\Sigma_{\gamma\gamma'}$) defined as
\begin{eqnarray}
\label{eq:R-self}
\Sigma_{\gamma\gamma'}(p,W)=\!\!\!\!\sum_{i}^{\pi\pi,K\bar{K}}\!\!\!
\sqrt{m_{R_\gamma}m_{R_{\gamma'}}\over E_{R_\gamma}(p)E_{R_{\gamma'}}(p)} 
\int_0^\infty\!\!\!
{E_i(q)\over \sqrt{E^2_i(q) + p^2}}
{q^2 f^{L_\gamma I_\gamma}_{R_\gamma,i}(q)
f^{L_{\gamma'}I_{\gamma'}}_{R_{\gamma'},i}(q)\ dq 
\over
W - E_\pi(p) - \sqrt{E^2_i(q) + p^2} + i\epsilon} \ .
\end{eqnarray}
The potential, $V^{J^PT}_{\alpha,\beta}(p',p; W)$, includes
Z-graphs [Fig.~1(a)] 
derived from the $\pi\pi$ model developed above.
(We ignore Z-graphs with $\pi K\bar{K}$
intermediate states for simplicity.)
\begin{figure}[t]
 \includegraphics[width=100mm]{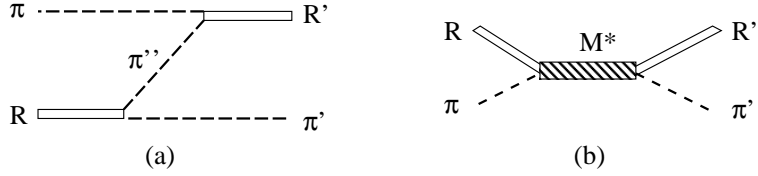}
\caption{$\pi$-$R$ interaction: (a) 3$\pi$ Z-graph; (b) $M^*$-term}
\end{figure}
The Z-graphs, together with the self-energy [Eq.~(\ref{eq:R-self})], are
essential to maintain the 3$\pi$ unitarity.
One may also include $M^*$-term [Fig.~1(b)] for the $\pi$-$R$ potential.

\section{Applications and results}

{\bf Effect of Z-graph on pole position}
We examine the effect of Z-graph on the pole position of $\pi_2(1670)$
and $\pi_2(2100)$. 
The procedure for finding the pole position suitable for our model has
been developed~\cite{suzuki}.
A pole position is at a complex energy ($W$) for which 
$W - M^0_{M^*} - \Sigma_{M^*}(W) = 0$ is satisfied; 
$M^0_{M^*}$ is the $M^*$ bare mass and 
$\Sigma_{M^*}(W)$ is the self energy of $M^*$ given by
\begin{eqnarray}
\label{eq:self-mstar}
\Sigma_{M^*}(W) = \sum_{\alpha,\beta}
\int^\infty_0\!\!\! q^2dq\ 
\bar{g}_{M^*,\alpha}(q)
G_{\alpha\beta}(q,W)
g_{M^*,\beta} (q) \ ,
\end{eqnarray}
where $g_{M^*,\beta}(q)$ is a $M^*\to \pi R$ vertex, and
$\bar{g}_{M^*,\alpha}(q)$ is a dressed vertex defined as
\begin{eqnarray}
\bar{g}_{M^*,\alpha}(p) =
g_{M^*,\alpha}(p) +
\sum_{\gamma\gamma'}
\int^\infty_0\!\!\! q^2dq\ g_{M^*,\gamma}(q)\ G_{\gamma\gamma'}(W,q)\
t^{J^PT}_{\gamma',\alpha} (q,p; W) \ ,
\end{eqnarray}
where $t^{J^PT}_{\beta,\alpha}$ is the T-matrix
calculated with Eq.~(\ref{eq:tmat}) in which the potential includes
only the Z-graphs.
We fit $M^0_{M^*}$ and $g_{M^*,\alpha}$ to pole
positions and branching ratios of $\pi_2$'s; data are taken from the
Particle Data Group.
Then we eliminate the Z-graphs from our model, 
that is, we calculate $\Sigma_{M^*}(W)$ with
Eq.~(\ref{eq:self-mstar}) in which the dressed vertex is replaced by the
bare one.
We find the pole
positions again from
$W - M^0_{M^*} - \Sigma_{M^*}(W) = 0$.
The pole for $\pi_2(1670)$ changes from 
$1672-130i$ MeV to $1689-133i$ MeV, and for $\pi_2(2100)$, 
from $2090-313i$ MeV to $2084-346i$ MeV.
Thus, the Z-graphs change the pole position by $\sim$ 10\% [$2\sim 3$\%] 
for $\pi_2(2100)$ [$\pi_2(1670)$].\\

\begin{figure}[t]
 \includegraphics[width=100mm]{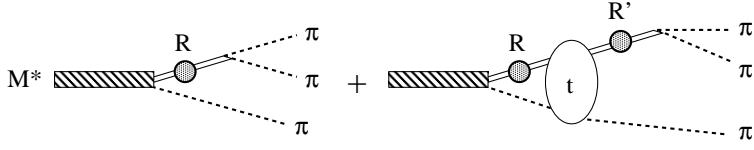}
\label{fig:mastar-decay}
\caption{$M^*$-decay amplitude: The bulb labeled 't' is the T-matrix
 calculated with Eq.~(\ref{eq:tmat}) in which the potential includes
 only the Z-graphs. The gray circle indicates the
 self-energy of the isobar [Eq.~(\ref{eq:R-self})].}
\end{figure}

\begin{figure}[t]
 \includegraphics[width=64mm]{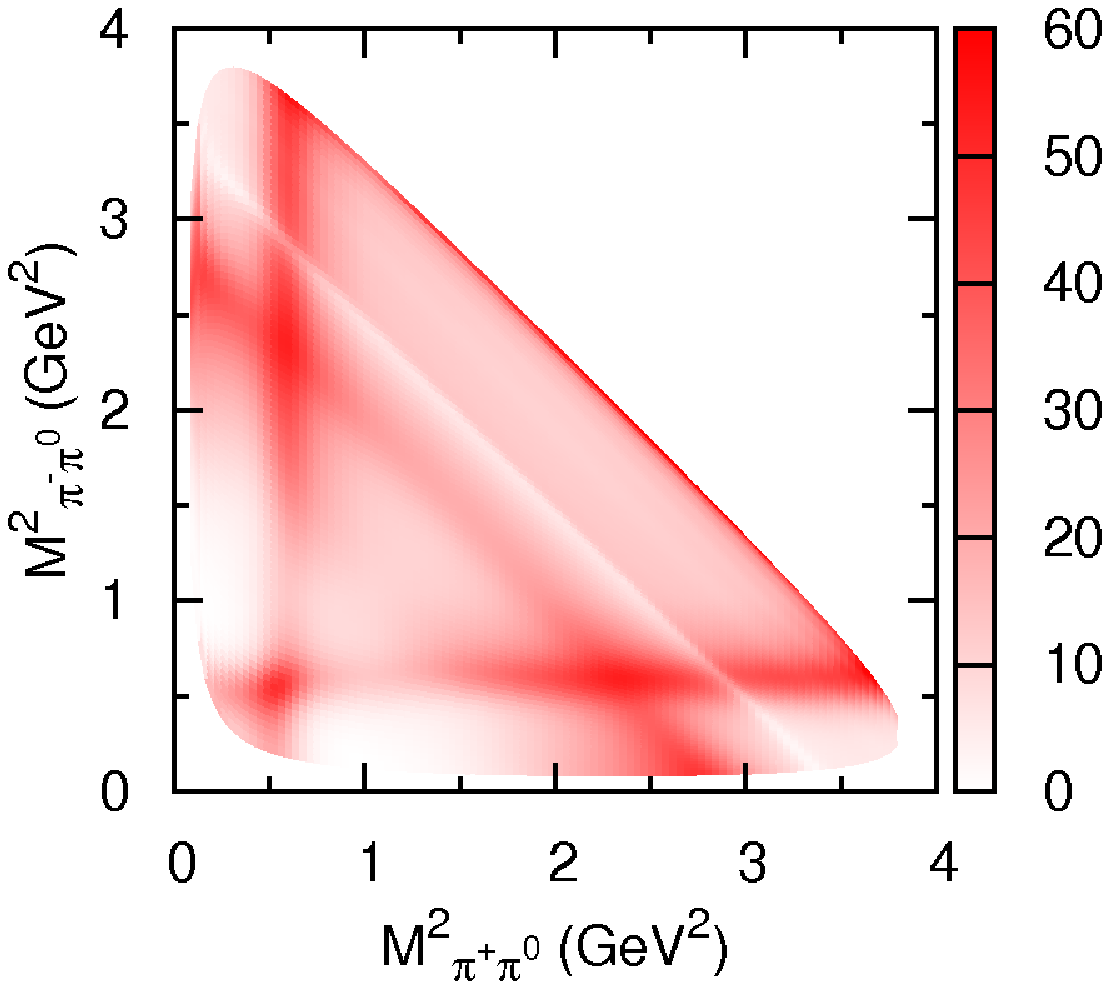}
\hspace{2mm}
 \includegraphics[width=65mm]{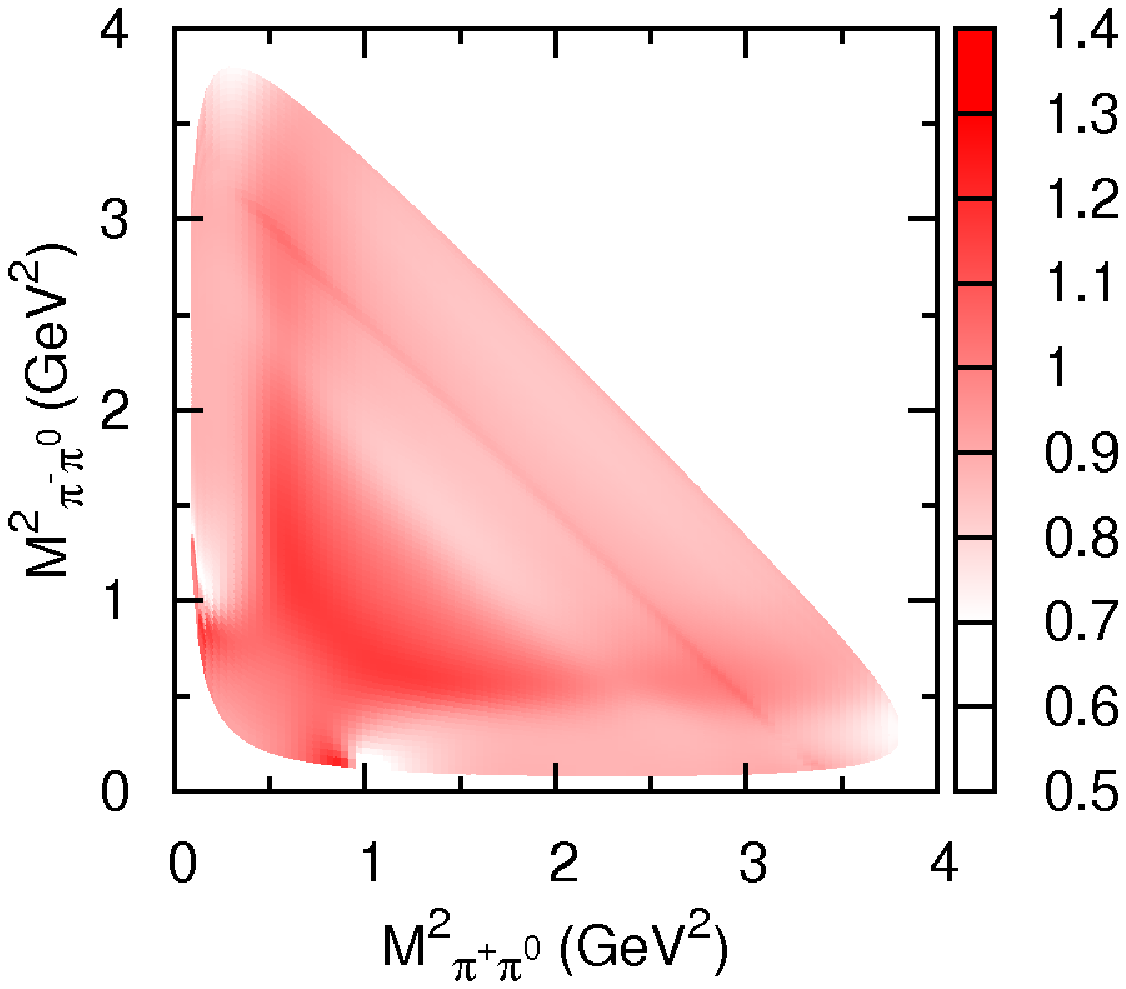}
\label{fig:dalitz}
\caption{(left) Dalitz plot of $\pi_2(2100)$ decay; (right) Ratio of
 Dalitz plots with and w/o the Z-graphs.}
\end{figure}

\noindent{\bf Effect of Z-graph on Dalitz plot}
We calculate the decay amplitude of a $M^*$ as graphically shown in
Fig.~\ref{fig:mastar-decay}, and then obtain the corresponding Dalitz plot.
In the first term of Fig.~\ref{fig:mastar-decay}, 
two pions are paired, and the 3rd pion is the
spectator. The second term includes rescatterings due to the Z-graphs.
Here, we show the Dalitz plot for the $\pi_2(2100)$ decay in
Fig.~\ref{fig:dalitz} (left). 
In order to examine the effect of the Z-graphs, we turn off the
Z-graphs, i.e., we calculate only the first term in Fig.~\ref{fig:mastar-decay}.
We show the ratio of the Dalitz plots with and without the Z-graphs in 
Fig.~\ref{fig:dalitz} (right). 
In general, the Z-graphs change both the magnitude and the shape of the
Dalitz plot. The effect is more significantly seen in the Dalitz plot than in
the shift of the pole position.
\begin{table}
\tabcolsep=6mm
\begin{tabular}{ccccc}\hline
  & $\Gamma_{\rm pole}$ (MeV)
  & $\pi_2\to\pi f_0$ (\%)
  & $\pi_2\to\pi \rho$ (\%)
  & $\pi_2\to\pi f_2$ (\%)\\
\hline
with Z & 626 & 45 & 19    & 36\\
w/o Z  & 542 & 36 & 27  & 37\\
\hline
\end{tabular}
\caption{The properties of $\pi_2(2100)$
extracted from models with (2nd row) and without (3rd row) the Z-graphs.
The second column is the width at the pole, 3-5th columns are the
 branching ratios.
}
\label{tab:pi2}
\end{table}

We may fit to the Dalitz plot of
Fig.~\ref{fig:dalitz} (left) with our model without the Z-graphs, by
varying the $M^*\to \pi R$ couplings and cutoff. 
With a tentative error of 5\% assigned to the Dalitz plot, we achieved
the fit with $\chi^2/{\rm data}\sim 1.1$.
It would be interesting to examine the properties of $M^*$ extracted
from the two models (with or without the Z-graphs) both of which
reproduce the same Dalitz plot.
In Table~\ref{tab:pi2}, we show the width and the branching ratios of
$\pi_2(2100)$  extracted from the two models.
We find a significant difference in the properties from the two models.
Although the effect of the Z-graphs depends on a system, still this results
indicates an importance of including the Z-graph (3$\pi$ unitarity) in
analyzing the Dalitz plot.



\begin{theacknowledgments}
The author thanks H. Kamano, T.-S. H. Lee and T. Sato for their
 collaboration at EBAC.
This work is supported by the U.S. Department of Energy, Office of
 Nuclear Physics Division, under 
 Contract No. DE-AC05-06OR23177 under which Jefferson Science Associates
 operates Jefferson Lab.
\end{theacknowledgments}



\bibliographystyle{aipproc}   


\IfFileExists{\jobname.bbl}{}
 {\typeout{}
  \typeout{******************************************}
  \typeout{** Please run "bibtex \jobname" to optain}
  \typeout{** the bibliography and then re-run LaTeX}
  \typeout{** twice to fix the references!}
  \typeout{******************************************}
  \typeout{}
 }


\end{document}